# Accurate Direct Measurement of Electrically Modulated Far-Field Thermal Infrared Emission and its Dynamics

Xiu Liu, Hakan Salihoglu, Xiao Luo, Hyeong Seok Yun, Lin Jing, Bowen Yu, Sheng Shen*

Department of Mechanical Engineering, Carnegie Mellon University, Pittsburgh, PA 15213, USA

*sshen1@cmu.edu

**Abstract**

Accurate direct measurements of far-field thermal infrared emission become increasingly important because conventional methods, relying on indirect assessments, such as reflectance/transmittance, are inaccurate or even unfeasible to characterize state-of-art devices with novel spectra, directionalities, and polarizations. The direct collection of the far-field emission from these tiny devices is also challenging because of their shrinking footprints and uncontrollable radiation noises from their surroundings. Here, we demonstrate a microscopic lock-in FTIR system that realizes significant improvement in signal-to-noise ratio (SNR) by combining a microscope and a lock-in amplifier with a Fourier transform infrared spectrometer (FTIR). The lock-in FTIR is ultrasensitive, with a specific detectivity $10^6$ times higher than commercial ones, to overcome the optical loss and background noise during the emission light collection. Based on an analytical model of the signal detection process, we also explore the combination of modulated Joule heating and global heating to fulfill the potential of our system for noise reduction. Our findings show that, compared to previous studies, more than 3 times lower temperatures are sufficient to generate a measurable signal. Under a heating temperature of around 125 °C, we can achieve a SNR of about 23.7, which is far above the true-signal-threshold (SNR of about 3.0). Furthermore, the system can respond fast enough (up to 175kHz) to record spectral-resolved dynamics of microdevices in the frequency domain. The measurable frequency range can be extended up to MHz or even GHz level by a high-speed circuit model. We believe the system together with the analytical signal processing can be beneficial for next-generation thermal infrared material and device exploration, boosting the applications in lighting, sensing, imaging, and energy harvesting on a small scale.

**Introduction**

Thermal emission from an object with a finite temperature can be exploited as a fingerprint in extracting material properties and used to measure temperature distributions based on Stefan-Boltzmann's law. Thermal radiation characteristics of bulk materials are typically isotropic, diffuse, and incoherent. With the advent of nanotechnology, thermal infrared microdevices enable versatile functionalities to control emission spectra[1–4], directionalities[5], and polarizations[6,7], thus opening new opportunities for applications in lighting, sensing, imaging, and energy harvesting[8–10]. The key for these applications is to develop accurate measurement techniques for characterizing thermal emission features of materials and devices. Conventional characterization techniques bear inherent limitations[11,12]. An indirect reflectance/transmittance method that relies on Kirchhoff's law for emissivity measurement can be unreliable for highly scattering or absorptive samples. Moreover, violation of Kirchhoff's law under non-equilibrium or non-reciprocal conditions leaves a direct emission measurement as the only method to obtain thermal emission properties. However, it becomes increasingly challenging to directly collect the weak radiation from miniatured and compact devices with small footprints, which is usually comparable with ambient thermal noises. Thus, a technique with a proper noise reduction is highly desired for increasing the signal-to-noise (SNR) in thermal emission measurements.

Several techniques have been employed to increase SNR. Intense global heating increases emission signals but gradually becomes impractical, since microdevices can be damaged at a high temperature or contain temperature-sensitive materials[13,14]. Coupling a lock-in amplifier (LIA) to a Fourier transform infrared spectrometer (FTIR) can filter out noises in near-field thermal radiation measurements with a modulated dithering tip[15–17]. The LIA-based noise reduction for far-field emission relies on driving a piezoelectric stage displacing the emitter periodically at 20 Hz[18]. Such a slow spatial modulation, however, still incurs the 1/f noise to some extent. A recent work directly measures emission spectra of a device based on a 10 kHz electrical modulation, which necessitates a full-wave rectified cosine wave of far-field emission with a

20 kHz reference frequency for LIA[19]. Nevertheless, the thorough analysis of modulated thermal infrared signals and their underlying physics are still lacking.

Achieving a fast modulation also allows us to characterize the dynamics of thermal emission. Recently, various thermal infrared devices have been employed to achieve rapid modulation of thermal emission with a speed up to GHz-level[19–26]. The typical method for characterizing the modulation speed of thermal emission is to measure the total emission power, which is not applicable for several modulation mechanisms, including peak shift for emission resonance tuning[3,21,22], spectrum variation for chemical sensing[27,28], etc. The technique to simultaneously monitor spectra and dynamics of thermal infrared emission is particularly beneficial for material and device exploration.

In this work, we demonstrate a microscopic lock-in FTIR system to directly measure the far-field thermal emission from an electrically modulated microdevice. To reduce the optical loss, a microscope is used to simultaneously locate the emission active region and extract the modulated thermal infrared signal to an FTIR. A LIA then demodulates the signal collected by the FTIR detector to significantly reduce optical and electrical noises during the measurement. We develop an equivalent thermal circuit model for the microdevice and exploit the model to elucidate the full signal detection process. We also employ the combination of the Joule and global heating to increase the SNR, resulting in a more than 3-fold reduction in heating temperatures to obtain a measurable signal comparable to the previous studies[13,14]. Under a heating temperature of around 125 °C, we can achieve a SNR of about 23.7, which is far above the true-signal-threshold (SNR of about 3.0)[29]. Furthermore, the microscopic lock-in FTIR enables fast response to a high modulation frequency.

**Method**

In our experimental schematics shown in Figure 1(a), a device under test (DUT) is globally heated up to a reference temperature $T_0$ by a heating stage and then electrically modulated by a voltage $V(t)$ controlled by an arbitrary wave generator (AWG; Keysight 33210A). The Joule heating, $Q(t)$, generated by the electrical modulation, changes the device temperature $T(t)$ over time, and the device then radiates the thermal infrared emission, $I(\lambda, T(t))$ (optical signal). An infinite-corrected reflective objective collects the modulated $I(\lambda, T(t))$ and directs it to the FTIR (Thermo Fisher iS50) for interferometry using a dichroic beamsplitter through an aperture. Here, the aperture can block the noises from structures surrounding the DUT and provide the space-resolved capability when the DUT emission is inhomogeneous. The beamsplitter transmits white light (from an external illumination) reflected by DUT to locate and focus on the active area of DUT using a microscope subsystem under a charge-coupled device (CCD) camera. The step-scan mode is used to avoid the double modulation issue arising when coupling the FTIR and the LIA[30]. The interferogram $S_{in}(\lambda, t)$ recorded by a Mercury-Cadmium-Telluride (MCT) detector is fed to the LIA (Zurich Instruments HF2LI), which works as a dynamic noise filter centered at the reference frequency that can be the modulation frequency or its higher order harmonics. The MCT detector that has an original specific detectivity of $8.0 \times 10^9$ Jones can be further boosted by the LIA with a dynamic reserve of 120dB, leading to a total specific detectivity of $8.0 \times 10^{15}$ Jones for our lock-in FTIR system. This ultrahigh specific detectivity allows us to overcome the heavy optical loss and background noise during the emission light collection. Finally, the output voltage $S_{out}(\lambda)$ from the LIA represents the emission intensity of the DUT.

To illustrate the signal detection process, we define the mathematical relations between the parameters in each processing step. The translation of the signal from the modulation voltage $V(t)$ to the thermal emission $I(\lambda, T(t))$ depends on a thermos-opto-electrical design of the DUT. In this work, we use an electrically driven thermal infrared metasurface with an active area (red-dashed box) of about $50 \times 50$ μm$^2$, as shown in Figure 1(b). The metasurface is composed of a gold nanorod array and adopts a center-contacted electrode line design, which not only enables a narrowband emissivity resonance at around 5.24μm, but also allows a fast Joule heating to modulate the thermal emission[31]. To obtain the temperature $T(t)$ of the metasurface under a given voltage $V(t)$, we neglect the top surface radiation and natural convection since heat dissipation into the substrate dominates. Since we only collect the emission from the active metasurface area (confined by the aperture) that is approximately isothermal, as verified by thermal infrared mapping[31], we can model the electrothermal design of the DUT as one-dimensional heat conduction to the substrate. To include the transient thermal response, we further simplify the heat transfer as an equivalent thermal circuit using the lumped capacitance model, as shown in Figure 1(c). Based on this model, we separate the substrate into two control volumes, represented by two nodes $T(t)$ and $T_0$, via the dashed line that is roughly defined by the thermal penetration depth. The control volume with $T(t)$ can store energy with a thermal capacitance $C_t$, or transfer heat to the control volume with $T_0$ (grounded) below via a thermal resistance $R_t$. The metasurface serves as a heat source $Q(t)$. Via the lumped capacitance model, we have the dynamic equation

$$C_t \frac{d(T(t) - T_0)}{dt} + \frac{T(t) - T_0}{R_t} = Q(t)$$

(1)

where $Q(t) = V^2(t)/R$ is from Joule heating and R is the electrical resistance of the metasurface. This linear first-order ordinary differential equation can be solved by constructing an exact differential with an integrating factor $\mu(t) = \frac{1}{C_t} e^{\frac{t}{R_t C_t}}$. The solution $T(t)$, with the initial condition $T(0) = T_0$, is given by

$$T(t) = T_0 + T_0 e^{-\frac{t}{R_t C_t}} + \frac{e^{-\frac{t}{R_t C_t}}}{C_t} \int_0^t e^{\frac{t}{R_t C_t}} Q(t) \, dt$$

(2)

With the derived temperature $T(t)$ of the emitting region, thermalization of which happens much faster than the temperature modulation speed, we can then compute the spectral thermal emission based on Planck's law in a time-pointwise manner as $I(\lambda, t) = \epsilon(\lambda) I_{BB}(\lambda, T(t))$ where $I_{BB}(\lambda, T(t))$ is the blackbody emission spectrum at temperature $T(t)$ and $\epsilon(\lambda)$ is the emissivity of the metasurface. The Joule heating generated by $V(t)$ has two parts: one is the time-independent part due to the root-mean-square of the power, and the other is the time-dependent part arising from modulation. Consequently, $T(t)$ can be expressed as $T(t) = T_0 + T_{DC} + T_m(t) = T_{avg} + T_m(t)$ where $T_{DC}$ and $T_m(t)$ are the time-independent and time-dependent parts of temperature, respectively. We can further group $T_0$ and $T_{DC}$ as the average temperature $T_{avg}$ of the emitting region, which can be measured from a thermal mapping system[31]. Since $T_m(t) \ll T_{avg}$, $I_{BB}(\lambda, T(t))$ can be expanded using the Taylor expansion with respect to $T(t)$ as

$$I(\lambda, t) = \varepsilon(\lambda)I_{BB}(\lambda, T_{avg}) + \varepsilon(\lambda)\frac{\partial I_{BB}}{\partial T}(\lambda, T_{avg})T_m(t)$$

(3)

which defines the thermal emission from the DUT (optical signal collected by the reflective objective). The optical signal reaching the MCT detector is then transformed to a voltage input $S_{in}(\lambda, t)$ to the LIA as

$$S_{in}(\lambda, t) = r(\lambda)I(\lambda, T) + S_{noise}$$
$$= r(\lambda)\left(\varepsilon(\lambda)I_{BB}(\lambda, T_{avg}) + \varepsilon(\lambda)\frac{\partial I_{BB}}{\partial T}(\lambda, T_{avg})T_m(t)\right) + S_{noise}$$

(4)

where $r(\lambda)$ is the response function whose wavelength dependence comes from MCT responsivity and other optical components along the optical path. $S_{noise}$ is the total unmodulated noise that can be filtered out by the LIA.

**Results**

Starting with the voltage input $S_{in}(\lambda, t)$ to the LIA in Eq.(4), we can explore the effects of Joule heating and global heating on the SNR. Specifically, we first demonstrate that the DC bias and AC waveshapes of Joule heating play a central role in optimizing the LIA for noise reduction. The reference temperature $T_0$ given by global heating, though unmodulated, is also found to significantly enhance the final output signal from the LIA.

To figure out the influence of the DC bias on the final detected signal, we start with a simple modulation voltage as

$$V(t) = V_p \cos(2\pi ft) + V_0$$

(5)

which is a cosine signal with a peak voltage $V_p$ and a DC bias $V_0$. The Joule heating can be expressed as $Q(t) = \frac{V^2(t)}{R} = Q_{DC} + Q_{1f}(t) + Q_{2f}(t)$ where the DC Joule heating is $Q_{DC} = \frac{V_p^2}{2R} + \frac{V_0^2}{R}$ and the first and second harmonics of Joule heating are $Q_{1f}(t) = \frac{2V_p V_0}{R} \frac{e^{i\omega t} + e^{-i\omega t}}{2}$ and $Q_{2f}(t) = \frac{V_p^2}{2R} \frac{e^{i2\omega t} + e^{-i2\omega t}}{2}$, respectively. Here, we use complex representations for convenience and keep the complex conjugate terms to fully express the real physical quantity since the measurement system is not all linear. The steady-state temperature of DUT is then given by inserting the DC and harmonic Joule heating terms into Eq.( 2 ), and we find that T(t) can also be expressed in terms of its harmonics as $T(t) = T_{DC} + T_{1f}(t) + T_{2f}(t)$ since T(t) is linearly dependent on Q(t). After substituting T(t) into Eq.( 4 ), we obtain the LIA input $S_{in}(\lambda, t)$. Based on the working principle of LIA, $S_{in}(\lambda, t)$ is then mixed with a reference $S_{ref}(t) = e^{-i(2\pi f_r t + \Delta\phi)}$ for a dual-phase demodulation, and the fixed phase difference $\Delta\phi$ between $S_{in}(\lambda, t)$ and $S_{ref}(t)$ can be finally cancelled[32]. The final output from the mixed signal $S_{mix}(\lambda, t) = S_{in}(\lambda, t) \cdot S_{ref}(t)$ depends on the order of its harmonics we choose via the reference frequency $f_r$. If we select $f_r = 1f$ (first harmonic), the magnitude of the DC output after the low-pass filter in the LIA is

$$S_{out,1f}^{cos}(\lambda) = r(\lambda)\varepsilon(\lambda) \frac{\partial I_{BB}}{\partial T}(\lambda, T_{avg}) \frac{V_p V_0 R_t}{R} \frac{1}{\sqrt{1 + (2\pi f)^2 R_t^2 C_t^2}}$$

( 6 )

and if selecting $f_r = 2f$ (second harmonic), we have the output signal

$$S_{out,2f}^{cos}(\lambda) = r(\lambda)\varepsilon(\lambda) \frac{\partial I_{BB}}{\partial T}(\lambda, T_{avg}) \frac{V_p^2 R_t}{4R} \frac{1}{\sqrt{1 + 4(2\pi f)^2 R_t^2 C_t^2}}$$

( 7 )

We also observe that the unmodulated portion of thermal emission together with the noises are always at $f_r \neq 0$, and finally filtered out.

Therefore, it is worth to note that we can use the first harmonic of the Joule heating for signal detection, unlike the previous studies[19] which always focus on the full-wave rectified cosine wave with 2f for reference frequency. More importantly, the filtered output signal $S_{out,1f}^{cos}(\lambda)$ is proportional to the DC bias $V_0$ such that we can overcome the power limit of the AWG by connecting a DC source in series to increase the signal input. To examine the dependence of the first and second harmonics of the output signal on $V_0$, we vary $V_0$ and keep constant $V_p = 1.0$ V and measure the emission peak $S^{cos}_{out,1f}(\lambda_r)$ at $\lambda_r = 5.24$ μm (at gray dashed line shown in Figure 2(b)) under $T_0 = 125$ °C and $f = 2600$Hz. For this measurement, we approximate $T_{avg} = T_{DC} + T_0 \approx T_0$ because $T_{DC}$ generated from the low modulation voltages is much smaller than $T_0$, and $2\pi f R_t C_t \approx 0$ because $T_0 \gg T_{DC}$ and $f \ll f_{3dB} = 1/2\pi R_t C_t$ where $f_{3dB}$ is the DUT cutoff frequency and can be verified in our frequency response measurements later. As shown in Figure 2(a), the emission peak $S^{cos}_{out,1f}(\lambda_r)$ increases linearly with increasing $V_0$ while the $S^{cos}_{out,2f}(\lambda_r)$ keeps approximately constant with fixed $V_p$. The crossing point of equal emission peak with about $V_p = 1.0$ V and $V_0 = 0.25$ V is well verified by theoretical predictions calculated using Eq.( 6 ) for $S^{cos}_{out,1f}(\lambda_r)$ and Eq.( 7 ) for $S^{cos}_{out,2f}(\lambda_r)$. These two formulas can also be verified by full spectra of the signals with two input voltage sets: ($V_p = V_0 = 1.0$ V) for $S^{cos}_{out,1f}(\lambda)$ and ($V_p = 2.0$ V, $V_0 = 0.0$ V) for $S^{cos}_{out,2f}(\lambda)$ in the LIA. Based on Eqs.( 6 ) and ( 7 ), they should have the same emission spectra, which are experimentally verified in Figure 2(b). Counterintuitively, the DC Joule heating contributes to the modulated signal based on its nonlinear electrical interference with the AC Joule heating as estimated in $Q_{1f}(t)$, which makes it fundamentally different from the global heating.

The output signal can also benefit from the waveshape of the modulation. If we change the cosine pulse train to a square pulse train, the modulation voltage can be defined as

$$V(t) = V_s D + \sum_{n=1}^{N=\infty} \frac{2V_s}{n\pi} \sin(nD\pi) \cos(n2\pi ft)$$

( 8 )

where D is the duty cycle, $V_s$ is the pulse height, and N denotes the number of harmonics. Following the same process as the above, we obtain the DUT temperature as $T(t) = T_0 + T_{DC} + T_{1f}(t) + \cdots$ and the magnitude of the $f_r = 1f$ (first harmonic) output voltage as

$$S_{out,1f}^{square}(\lambda) = r(\lambda)\varepsilon(\lambda)\frac{\partial I_{BB}}{\partial T}(\lambda, T_{avg})\frac{R_t}{2R}\left\{\frac{4V_s^2 D}{\pi}\sin(D\pi) + \frac{V_s^2}{\pi^2}\sin(D\pi)\sin(2D\pi)\right.$$

$$+\frac{2V_s^2}{\pi^2}\sum_{n=2}^{N=\infty}\left[\frac{1}{n(n-1)}\sin(nD\pi)\sin((n-1)D\pi)\right.$$

$$\left.\left.+\frac{1}{n(n+1)}\sin(nD\pi)\sin((n+1)D\pi)\right]\right\}\frac{1}{\sqrt{1+(2\pi f)^2 R_t^2 C_t^2}}$$

(9)

We compare the square pulse train ($V_s = 2.0$ V, $D = 50$ %) and the cosine pulse train ($V_p = V_0 = 1.0$ V) under $T_0 = 125$ °C and $f = 2600$ Hz, where $V_s$, $D$, $V_0$, and $V_p$ are set to have the same period and pulse height for both pulse trains except for the waveshape. $T_0$ and $f$ are set to maintain the approximations of $T_{avg} \approx T_0$ and $2\pi f R_t C_t \approx 0$, respectively. A $4/\pi$ factor increment in the output signal is expected and is observed in Figure 2(b) with the emission peak ratio $S_{out,1f}^{square}(\lambda_r)/S_{out,1f}^{cos}(\lambda_r)$ at emission resonance $\lambda_r = 5.24$ μm (at gray dashed line) about 1.2. In addition, $S_{out,1f}^{square}(\lambda)$ is also a function of D. To demonstrate this, we change D and measure the emission peak $S_{out,1f}^{square}(\lambda_r)$ with $V_s = 3.0$ V, $T_0 = 125$ °C and $f = 2600$ Hz, whose data is well consistent with the analytical expression $S_{out,1f}^{square}(\lambda_r, D)$ from Eq.( 9 ). Here, we normalize $S_{out,1f}^{square}(\lambda_r)$ and $T_{avg}$ to their values with $D = 50\%$, respectively, and use $N = 13$ in our computation because the number of harmonic terms required to approximate a square pulse waveshape depends on the DUT response speed. More specifically, $Nf \leq f_{3dB}$ where $f_{3dB} = 33.275$ kHz is the cutoff frequency from our dynamic measurements (discussed below). More advanced waveshapes of the modulation can be applied to further benefit the emission measurement, all of which can be accurately explored based on our signal process analysis.

After deriving the formulas for the LIA output (Eqs.( 6 ), ( 7 ), and ( 9 )), we are now able to demonstrate the effect of global heating on the SNR. We choose a square pulse train with $V_s = 3.0$ V, $D = 50\%$, and $f = 2600$ Hz as the modulation voltage V(t) and measure the emission spectra $S_{out,1f}^{square}(\lambda)$ under different global heating temperatures $T_0$. As shown in Figure 3(a), with $T_0$ increasing from 25°C to 150°C, the intensity of $S_{out,1f}^{square}(\lambda)$ enhances significantly because $T_0$ greatly increases the blackbody emission as indicated by the term $\partial I_{BB}(\lambda, T_{avg})/\partial T$ in Eq.( 9 ), although $T_0$ itself is unmodulated. This increasing effect is strong and overwhelms the decreasing effect ($S_{out,1f}^{square}(\lambda) \propto 1/R$ from Eq.( 9 )) from the electrical resistance R that is increased by the global heating temperature $T_0 \approx T_{avg}$. Since the influence from blackbody emission $\partial I_{BB}(\lambda, T_{avg})/\partial T$ is the same for all the harmonics of $S_{out,1f}^{square}(\lambda)$, the emission enhancement from $T_0$ works for any waveshape for Joule heating, including the simple cosine waveshape as indicated by Eqs.( 6 ) and ( 7 ). To specify the SNR increment from $T_0$, we repeatedly measure the emission spectra of $T_0 = 25\ °C$ and $125\ °C$ for 5 times and calculate the SNR based on[33]

$$SNR(\lambda) = \frac{\overline{S}_{out,1f}^{square}(\lambda) - S_{out,1f}^{Off}(\lambda)}{\sigma_{out,1f}^{square}(\lambda)}$$

( 10 )

where $\overline{S}_{out,1f}^{square}(\lambda)$ and $\sigma_{out,1f}^{square}(\lambda)$ are the mean and standard deviation of repeatedly measured emission spectra, respectively. $S_{out,1f}^{Off}(\lambda)$ is the emission spectrum when the DUT is off, i.e., without Joule heating. As shown in Figure 3(b), even under the room temperature (25 °C), the SNR(λ) can go beyond the true-signal-threshold[29] (SNR = 3.0; red-dashed line) near the resonance region with $SNR(\lambda_r) = 7.9$. The region that is above the true-signal-threshold then extends notably when we increase $T_0$ to 125 °C, and $SNR(\lambda_r)$ reaches to 23.7. Thanks to the strong noise reduction from our lock-in FTIR system, we only require a much smaller $T_0$ (25~125 °C) compared to previous direct measurements of far-field emission that is generated by a temperature more than 480 °C[13,14]. The heating temperature required by our system is only about 1/3 or even 1/20 of that needed in the previous studies, and this temperature can be further reduced if we keep optimizing the parameters setup in the LIA. We also notice that the SNR drops at the wavelengths around 3~4 μm and 6~7 μm due to the optical loss from $H_2O$ and $CO_2$ absorption, respectively, which can be solved by purging our optical paths with the $N_2$ gas.

After obtaining the emission spectra with a high SNR, we can further characterize the frequency response of the DUT. Under $T_0 = 125\ °C$, we choose the square pulse train with $V_s = 3.0\ V$ and $D = 50\%$ to take advantage of the $4/\pi$ increment factor as compared to a conventional cosine pulse train. At the same time, the 50% duty cycle enables to get rid of the summation of N harmonic terms from the waveshape deformation that can complicate the frequency response

measurement, as discussed in Figure 2(c). The measured output voltage is then obtained from Eq.( 9 ) as

$$S_{out,1f}^{square}(\lambda, f) = r(\lambda)\varepsilon(\lambda)\frac{\partial I_{BB}}{\partial T}(\lambda, T_{avg})\frac{V_s^2 R_t}{\pi R}\frac{1}{\sqrt{1 + (2\pi f)^2 R_t^2 C_t^2}}$$

( 11 )

We record the entire emission spectrum at multiple modulation frequencies f and characterize the dynamic response with f up to the DUT cutoff frequency $f_{3dB}$ as shown in Figure 4(a). With the curve fitting by Eq.( 12 ) in Figure 4(b), we can obtain the thermal time constant $R_t C_t = 4.78$ μs and the corresponding cutoff frequency as $f_{3dB} = 33.275$ kHz. It is noteworthy that we can obtain the full spectrum of the emission as our dynamic response characterization is spectrally resolved. This can be particularly useful for the modulation mechanisms other than the emission power.

The measurable frequency range of our system is up to 175 kHz, which is limited by the preamplifier connecting to the MCT detector. With an acceptable reduction of the specific detectivity, it is possible to remove this preamplifier to fully use the bandwidth of the MCT detector (expected about 20 MHz), which is enough for the fastest ever reported temperature-modulated thermal infrared devices[19]. The LIA bandwidth (50 MHz) is not a limiting factor. The time constant of the low-pass filter, typically long for noise reduction, has nothing to do with this frequency-domain measurement since the signal at this stage is at 0 Hz after mixing. For devices with emissivity modulation by a voltage or a current, we can further extend the frequency range to the GHz level based on a high-speed circuit model[22,34] with an electrical resistance connected to the device in series.

The electrothermal model of our metasurface is currently lumped as an equivalent thermal circuit model, and it is possible to include the spatial thermal analysis, such as the lateral temperature distribution and vertical thermal penetration depth. With the spectral-resolved dynamic thermal emission, we can inversely obtain key information in material properties and device performances of the thermal infrared microdevices.

**Conclusion**

In summary, with a combination of a microscope, LIA, and FTIR, we propose and demonstrate a measurement technique, microscopic lock-in FTIR, to directly measure the far-field emission of an electrothermal metasurface. The lock-in FTIR is ultrasensitive with a specific detectivity of $8.0 \times 10^{15}$ Jones allowing to overcome the heavy optical loss and background noise during the emission light collection. The microscope subsystem can further reduce the optical loss by a proper optical path alignment together with a $N_2$ purging. With an analysis of the nonlinear signal detection processing from the initial heat generation to the final measured signal output, our method exploits the combination of global and modulated Joule heating for the SNR improvement, allowing a more than 3 times lower temperature to obtain a measurable signal compared to the previous studies[13,14]. Under a heating temperature of around 125 °C, we can achieve a SNR of about 23.7, which is far above the true-signal-threshold. Furthermore, our system can be applied for a spectral-resolved frequency-domain characterization of the device response and can monitor the entire spectra for each modulation frequency providing more key information about the device performance. The measurable frequency range can be extended to MHz or even GHz with simple adjustments. We expect that our microscopic lock-in FTIR system will play a significant role in the characterization of modern thermal infrared microdevices.

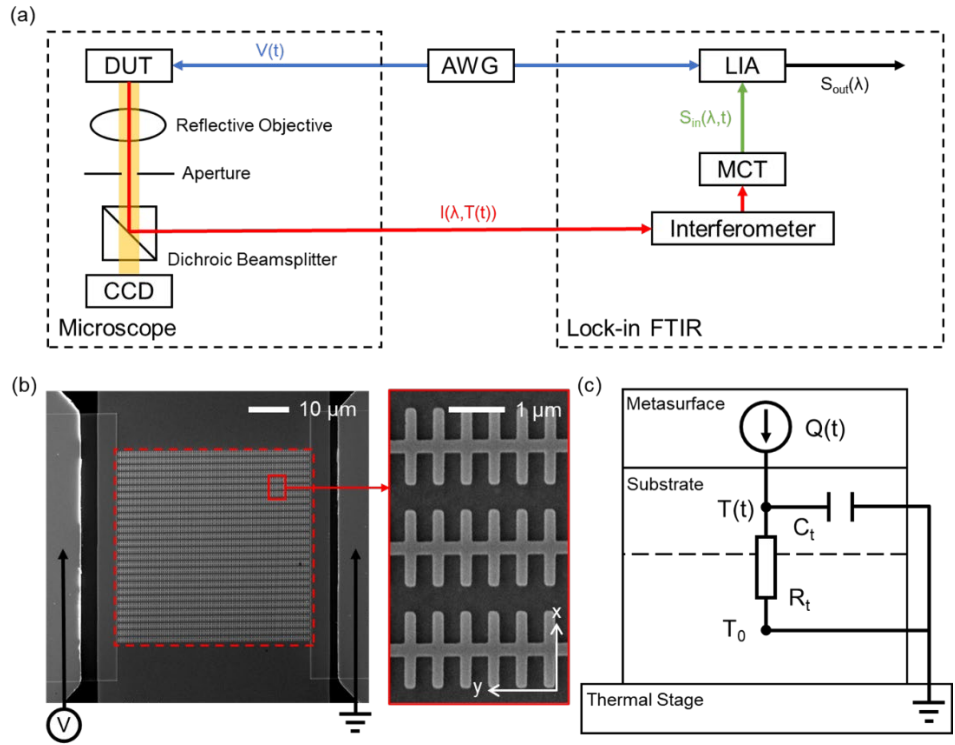

Figure 1: Measurement setup and device design. (a) Schematics of the microscopic lock-in FTIR system setup and its signal detection process. (b) Top view of SEM images of the electrothermal metasurface with a gold nanorod array for narrowband emission. The emission active area (red-dashed box) is about $50 \times 50 \ \mu m^2$. The inset shows the dimensions of the metasurface: periodicities $P_x = 1.8 \ \mu m$, $P_y = 0.5 \ \mu m$, length $L = 1.3 \ \mu m$, and width $W = 0.15 \ \mu m$. (c) Schematics of an equivalent thermal circuit model for the electrothermal response of the metasurface.

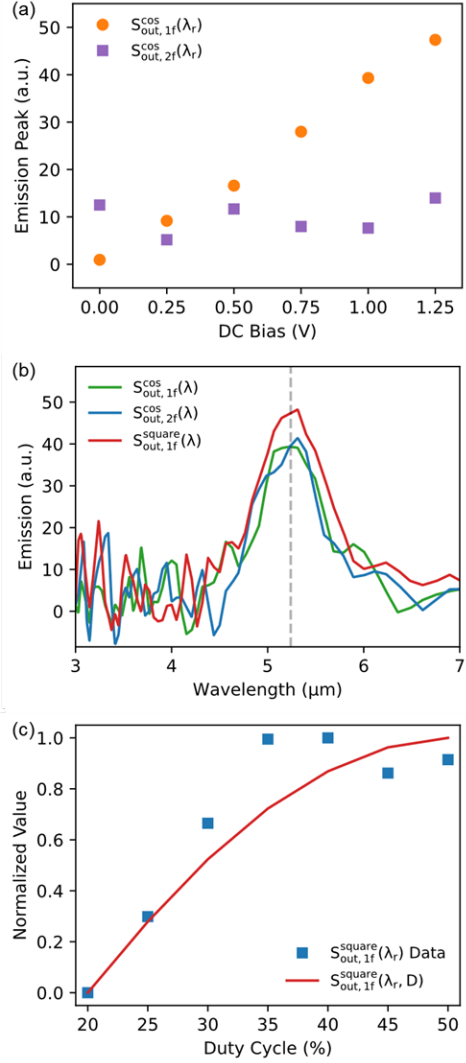

Figure 2: Analysis of the effect of Joule heating on thermal emission. (a) Emission peaks $S^{cos}_{out,1f}(\lambda_r)$ and $S^{cos}_{out,2f}(\lambda_r)$ at $\lambda_r = 5.24$ μm from the LIA under different DC biases $V_0$, under $V_p = 1.0$ V, $T_0 = 125$ °C, and $f = 2600$Hz. (b) Emission spectra of $S^{cos}_{out,1f}(\lambda)$ with ($V_0 = V_p = 1.0$ V) and $S^{cos}_{out,2f}(\lambda)$ with ($V_p = 2.0$ V, $V_0 = 0.0$ V), under $T_0 = 125$ °C and $f = 2600$Hz. The waveshape can further boost the SNR, as indicated by the $4/\pi$ factor increment at resonance $\lambda_r$ (gray-dashed line) of $S^{square}_{out,1f}(\lambda)$ by a square pulse train modulation ($V_s = 2.0$ V, $D = 50\%$) compared to $S^{cos}_{out,1f}(\lambda)$. (c) Normalized emission peak data $S^{square}_{out,1f}(\lambda_r)$, as a function of duty cycle D of the square pulse train, under $V_s = 3.0$ V, $T_0 = 125$ °C, and $f = 2600$Hz, keeps consistent with the analytical formula $S^{square}_{out,1f}(\lambda_r, D)$.

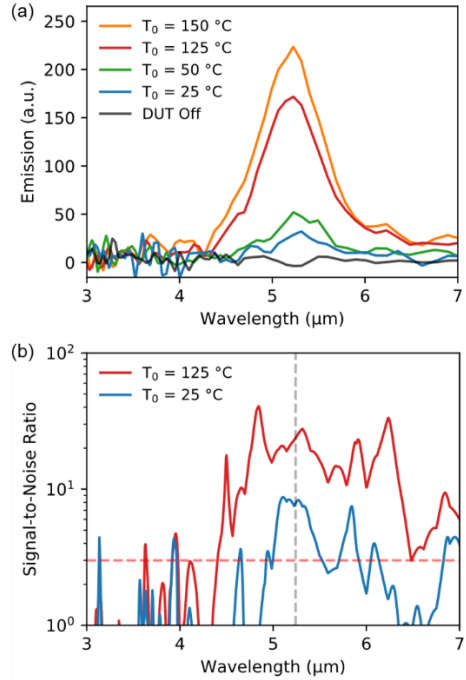

Figure 3: Analysis of the effect of global heating on thermal emission. (a) The emission spectra $S_{out,1f}^{square}(\lambda)$ under different $T_0$, under $V_s = 3.0$ V, $D = 50\%$, and $f = 2600$ Hz. Although unmodulated, $T_0$ increases the blackbody emission resulting in the magnitude increment for $S_{out,1f}^{square}(\lambda)$. (b) The SNR for $T_0 = 25$ °C and $125$ °C with 5 times repeated measurements. With $T_0$ increasing from 25 °C to 125 °C, the region above the true-signal-threshold (SNR about 3.0; red-dashed line) enlarges with SNR at resonance $\lambda_r = 5.24$ μm (gray-dashed line) increases from 7.9 to 23.7.

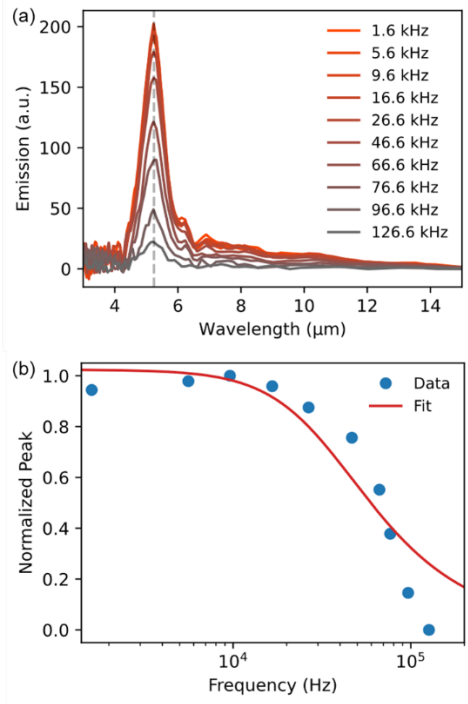

Figure 4: Spectral-resolved frequency-domain dynamic characterization (a) Emission spectrum $S_{out,1f}^{square}(\lambda)$ at each modulation frequency f, under a square pulse train modulation V(t) ($V_s$ = 3.0 V, D = 50%) and $T_0$ = 125 °C. (b) Emission peaks $S_{out,1f}^{square}(\lambda_r)$ at $\lambda_r$ = 5.24 μm under different modulation frequencies f. A curve fitting based on the analytical model $S_{out,1f}^{square}(\lambda_r, f)$ can obtain the device thermal time constant $R_t C_t$ as 4.78 μs.